\documentclass{aa}

\usepackage{graphicx}

\newcommand{\reduceme}{\mbox{R\kern-.10em\lower.35ex\hbox{E}\kern-.10em%
\hbox{D}\kern-.12em\raise.85ex\hbox{uc}\kern-.90em\lower.35ex\hbox{m}E}}
\newcommand{\xx}[1]{\makebox[0in][l]{$^{#1}$}}
\newcommand{\z}{[Fe/H]}
 
\begin{document}

\thesaurus{(04.03.1;   
            08.06.3;   
            11.19.5)}  

\title{Empirical calibration of the $\lambda$4000~\AA\ break}

\author{J.\,Gorgas\inst{1}             
   \and N.\,Cardiel\inst{1}            
   \and S.\,Pedraz\inst{1}             
   \and J.\,J.\,Gonz\'{a}lez\inst{2}} 

\offprints{J. Gorgas}
\mail{fjg@astrax.fis.ucm.es}

\institute{Departamento de Astrof\'{\i}sica, Facultad de Ciencias F\'{\i}sicas,
Universidad Complutense de Madrid, 28040 Madrid, Spain
\and
Instituto de Astronom\'{\i}a, U.N.A.M., Apdo. Postal 70-264, 04510 M\'{e}xico
D. F., M\'{e}xico}

\date{Received date / Accepted date}

\maketitle

\begin{abstract}
Empirical fitting functions, describing the behaviour of the
$\lambda$4000~\AA\ break, D$_{4000}$, in terms of effective temperature,
metallicity and surface gravity, are presented. For this purpose, the break
has been measured in 392 stars from the Lick/IDS Library. We have followed a
very detailed error treatment in the reduction and fitting procedures,
allowing for a reliable estimation of the break uncertainties. This
calibration can be easily incorporated into stellar population models to
provide accurate predictions of the break amplitude for, relatively old,
composite systems.
\keywords{Catalogs -- Stars: fundamental parameters -- 
Galaxies: stellar content} \end{abstract}

\section{Introduction}

The spectroscopic study of the blue and near-UV region around
$\lambda$4000~\AA\ has proven to be a useful tool to investigate the stellar
populations of composite stellar systems.  Obviously, this spectral region is
specially suited to detect the presence of young stars and therefore, to study
star formation histories. Furthermore, in a pioneering work,
Morgan (1959) already showed that the difference in intensity of the continuum
level on the two sides of H$\zeta$ ($\lambda3889$\AA) in galactic globular
clusters correlated well with metal abundance.

Although different absorption line-strength indices have been defined in this
spectral range to understand the stellar composition of early-type objects
(e.g., Faber 1973; Burstein et al. 1984; Rose 1984, 1994; Tripicco 1989; 
Wor\-they et al. 1994, hereafter W94; Jones \& Worthey 1995; 
Worthey \& Ottaviani 1997; Vazdekis \&
Arimoto 1999), some of them are quite dependent on spectral resolution (and
therefore velocity dispersion) and, in many cases, their use requires
relatively high signal-to-noise ratios. In this sense, an interesting spectral
index that avoids these problems is the $\lambda$4000~\AA\ break. We have
already demonstrated (Cardiel et al. 1998) that this discontinuity can be
measured with a relative error of $\sim 10$\% with a signal-to-noise ratio per
\AA\ $\sim 1$.  Thus, the break is well suited to be measured in faint objects
or at low surface brightnesses. However, this advantage, due to the large
wavelength interval employed in its definition, also translates into an
important drawback: many absorption lines are included in the break
bandpasses. Therefore, the behaviour of the break is expected to be
complex. 

In this work we are using the definition adopted by Bruzual (1983), who
defined this spectral index as the ratio of the average flux density, $F_\nu$
(erg s$^{-1}$ cm$^{-2}$ Hz$^{-1}$), in two bands at the long--
and short--wavelength side of the $\lambda$4000~\AA\ discontinuity, in
particular
\begin{equation}
{\rm D}_{4000} =
  \frac{(\lambda_2^{-}-\lambda_1^{-})}{(\lambda_2^{+}-\lambda_1^{+})}
  \frac{{\displaystyle \int_{\lambda_1^{+}}^{\lambda_2^{+}}}
                            F_\nu \; {\rm d}\lambda}%
       {{\displaystyle \int_{\lambda_1^{-}}^{\lambda_2^{-}}}
                            F_\nu \; {\rm d}\lambda},
\label{definitionD4000}
\end{equation}
where \mbox{$(\lambda_1^{-},\lambda_2^{-},\lambda_1^{+},\lambda_2^{+})
= (3750,3950,4050,4250)$~\AA}. Except for the combination of $\nu$ and
$\lambda$, (due to the measurement method employed by Bruzual, in
which the break was obtained from galaxy spectra plotted as $F_\nu$
versus $\lambda$), and for not being in logarithmic units, this
definition resembles that of a color. Because of this, the D$_{4000}$
can be considered as a pseudo-color. However, it must be pointed out
that, compared to the classical color indices in this spectral range
(like ${\rm U}-{\rm B}$), the 4000~\AA\ break is much less sensitive
to extinction by dust, and hence, it is more valuable to investigate
the stellar content of galaxies. In particular, using the average
interstellar extinction curve from Savage \& Mathis (1979), the D$_{4000}$
can be corrected from internal reddening (and galactic extinction for
objects at zero redshift) using the following expression
\begin{equation}
{\rm D}_{4000}^{\rm corrected}={\rm D}_{4000}^{\rm observed}\;
10^{-0.0988\; {\rm E(B-V)}}.
\end{equation}
It is
interesting to note that the break definition is still the same for redshifted
objects, where the integral limits must be properly modified, and that absolute
fluxes are not required. As a working expression, the 4000~\AA\ break can be
rewritten as
\begin{equation}
{\rm D}_{4000} =
  \frac{{\displaystyle \int_{4250\,(1+z)}^{4050\,(1+z)}}
                            \lambda^2 f_\lambda \; {\rm d}\lambda}%
       {{\displaystyle \int_{3750\,(1+z)}^{3950\,(1+z)}}
                            \lambda^2 f_\lambda \; {\rm d}\lambda},
\label{definitionD4000bis}
\end{equation}
where $f_\lambda$ is the flux density per unit wavelength (in arbitrary units),
and $z$~is the redshift of the object being measured.

Up to date, many authors have employed the D$_{4000}$ to study the stellar
composition and star formation history of early-type galaxies (e.g. McClure \&
van den Bergh 1968; Spinrad 1980, 1986; Bruzual 1983; Laurikainen \& Jaakkola
1985; Hamilton 1985; Dressler 1987; Dressler \& Shectman 1987; Johnstone et
al. 1987; Kimble et al. 1989; Dressler
\& Gunn 1990; Rakos et al. 1991; Charlot et al. 1993; Charlot \& Silk 1994;
Davidge \& Clark 1994; Songaila et al. 1994; Belloni et al. 1995; Davidge \&
Grinder 1995; Cardiel et al. 1995, 1998; Abraham et al. 1996; Hammer et al.
1997; Barbaro \& Poggianti 1997; Longhetti et al. 1998; Ponder et al. 1998).
The reliable analysis of the break measurements rests on the comparison of the
data with the predictions of stellar population models (e.g. Worthey 1994;
Bruzual \& Charlot 1996). So far, such predictions are computed by using either
model atmospheres, or stellar libraries with a poor coverage of the
atmospheric parameter space, especially in metallicity.

In this paper we present an empirical calibration of the $\lambda$4000~\AA\
break as a function of the main atmospheric stellar parameters (namely
effective temperature, surface gravity and metallicity) in an ample stellar
library which covers an appropriate range of parameters to study relatively old
stellar populations.  One of the main advantages of using fitting functions to
describe the behaviour of spectral indices is that they allow stellar
population models to include the contribution of all the required stars,
through a smooth interpolation in the space defined by the fitted stellar
parameters. The usefulness of this approach has been demonstrated by the
successful inclusion of similar fitting functions in recent evolutionary
synthesis models (eg. Worthey 1994; Vazdekis et al. 1996; Bressan et al. 1996;
Bruzual \& Charlot 1996). 

It is important to keep in mind that the empirical
calibration is only a mathematical representation of the break behaviour 
as a function of atmospheric stellar parameters, and that
we do not attempt to obtain any physical justification of the derived
coefficients.

We briefly review the previous works devoted to understand the D$_{4000}$ in
section~2. The star sample is given in section~3. The observations and data
reduction are described in section~4, whereas section~5 contains a description
of the error analysis. In section 6 we show the behaviour of the measured
D$_{4000}$ values as a function of the stellar atmospheric parameters. The
fitting strategy and the resulting empirical function are presented in
sections~7. Finally, in section~8 we give a summary,  
providing a public FORTRAN subroutine written by the authors to 
facilitate the
computation of the D$_{4000}$ using the fitting function presented in this
paper. Sections~4 and~5 are rather technical, due to the inclusion of a
lengthy explanation of the data and error handling. We suggest the reader
not interested in such details to scan Tables~\ref{tabletotal}
and~\ref{tableruns}, and skip those sections.

\section{Previous works: understanding the D$_{4000}$}

\begin{figure}
 \begin{center}
  \resizebox{\hsize}{!}{\includegraphics[angle=-90]{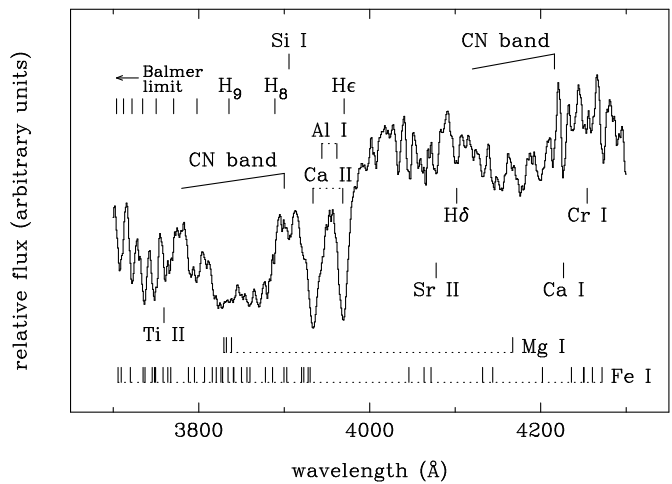}}
  \caption{Spectrum of the star HD72324 (G9~III) in the region around the
$\lambda$4000~\AA\ break. As a reference, we have also plotted the most intense
(${\rm EW} > 200$~m\AA) Fraunhofer lines from the sun (using the
tabulated data from Lang 1974 ---original source Moore et al. 1966---),
together with the Balmer lines and two CN molecular bands (the central bandpass
of the CN3883 index defined by Pickles 1985, and the absorption band of
the S(4142) index employed by Smith et al. 1997) which can be found in
this spectral range. The contribution of the atomic metallic lines, especially
from Fe~I and Mg~I, becomes very important bluewards $\lambda$4000~\AA.}
  \label{figd4000detailed}
 \end{center}
\end{figure}

The $\lambda$4000~\AA\ break is a sudden onset of absorption features
bluewards 4000~\AA\ which is clearly noticeable for stellar types
cooler than G0 (see Fig.~\ref{plotspt}). In Fig.~\ref{figd4000detailed} we
show a typical spectrum of a cool star in this spectral region, together with
the identification of the most prominent spectral features. Considering the
large wavelength range employed in the measurement of the D$_{4000}$, it is
expected the strength of this discontinuity to be a function of the
distribution of the continuum light in this region (governed by the effective
temperature) modulated by the absorption line strengths (which must depend
primarily on both temperature and metallicity, and secondly on gravity).  This
behaviour converts the break in a potential tool to investigate composite
stellar populations in early-type systems.

The relevance of the line-blanketing discontinuity near $\lambda$4000~\AA\ was
the object of a systematic study by Wildey et al. (1962). These authors
measured the {\it energy subtracted\/} in the spectra of some stars due to
Fraunhofer lines, showing that the effect was important below
$\lambda$4000~\AA.  Van den Bergh (1963), and van den Bergh \& Sackmann (1965)
defined a break, $\Delta$, as the ratio of the smoothed observed continuum at
both sides of $\lambda$4000~\AA. These authors measured this break in a sample
of 200~stars concluding that $\Delta$ depended both on stellar metallicity and
B$-$V color. Analogous discontinuity definitions, like \mbox{$C(38-41)$}
(McClure \& van den Bergh 1968), and
\mbox{$\Gamma(38-41)$} (Carbon et al. 1982), have been also employed in the
spectroscopic analysis of stars, star clusters and galaxies.

Using spectrophotometric stellar libraries, Bruzual (1983) and 
Hamilton (1985)
studied the variation of the $\lambda$4000~\AA\ break with spectral types and
luminosity classes (compare Fig.~3 in Bruzual with Fig.~6 in Hamilton). Both
authors concluded that, as a function of temperature, the D$_{4000}$ increases
slowly for spectral types in the range from O5 to G0, and faster from G0 to
M0, whereas the break decreases for the later types, M0 to M5. In addition,
whilst for spectral types hotter than G0 the break does not depend on gravity,
a clear dichotomy between main sequence stars on one hand, and giant and
supergiant stars on the other, is apparent for lower temperatures. Given the
scarcity of the employed stellar libraries, no dependence on metallicity could
be obtained in these works.

From the analysis of moderate-resolution spectra of 950 galaxies in 12 rich
clusters, Dressler \& Shectman (1987) argued that, in composite stellar
populations, the break is insensitive to changes in metal abundance, at least
in the metallicity range spanned by their galaxy sample. This result was
employed by Munn (1992) to conclude that the CN--D$_{4000}$ diagram is
effective at separating metallicity and age effects on the integrated spectra
of early-type galaxies. However, Kimble et al. (1989) obtained that the break
correlated strongly with metallicity indicators, such as the Mgb index.

More recently, Poggianti \& Barbaro (1997), working with Kurucz's models, have
obtained a theoretical calibration of the break as a function of stellar
parameters. They present (Fig.~1 in their
paper) the behaviour of the D$_{4000}$ in the ranges \mbox{$5500 < T_{\rm eff}
< 35000$~K}, \mbox{$0 < \log g < 5$}, and \mbox{$-2 < \log Z < 0$}.  This work
clearly shows the strong dependence of the break on effective temperature, as
previously reported from the studies based on stellar libraries, and
quantifies, for a small sample of temperatures, the variation of the break as a
function of metallicity and gravity. The D$_{4000}$ is shown to be insensitive
to metallicity for hot stars ($T_{\rm eff} = 9000$~K), whereas the
contrary is true for \mbox{$T_{\rm eff} = 5500$}~K. In
addition, using the stellar spectra of Straizys \& Sviderskiene (1972; note
that these spectra are those also employed by Bruzual 1983), these authors
obtain that, for stars with \mbox{$3500 < T_{\rm eff} < 5500$~K}, the
D$_{4000}$  always exhibits values above 2, with a maximum of 3 at
\mbox{$T_{\rm eff}=4000$~K}.
Using this theoretical calibration, Barbaro \& Poggianti (1997) have also
elaborated an evolutionary synthesis model which predicts, in the integrated
spectrum of a galaxy, the variation of the D$_{4000}$ as a function of the
star formation rate (SFR). More interestingly, they conclude that the break
can be employed to yield the ratio of the SFR averaged over the last 5 billion
years to the present SFR.

From all these previous works, it is quite clear that the D$_{4000}$ is a
suitable tool to face the study of stellar systems, in particular to reveal
their stellar composition. However, a detailed empirical calibration, such as
that presented in this paper, is needed to i) overcome the unavoidable
uncertainties associated to the theoretical calibrations, ii) extend our
understanding of the break behaviour for stars with $T_{\rm eff} < 5500$ K
(note that these late-type stars constitute a fundamental ingredient in the
modeling of old stellar populations), and iii) use in
conjunction with other indices previously calibrated with the same stellar
library. 

\section{Star sample}

In order to derive a confident empirical calibration of the D$_{4000}$, we
decided to measure this spectral feature in all the stars belonging to the
Lick/IDS Library (Burstein et al. 1984; Faber et al. 1985;
Burstein et al. 1986; Gorgas et al. 1993, hereafter G93; and Worthey et
al. 1994). The suitability of the Lick/IDS library to
obtain empirical fitting functions has been widely demonstrated by the works
of G93, W94, and Worthey \& Ottaviani (1997), who, in overall, have derived
analytical expressions for 25 spectral indices in the \mbox{4000--6000}~\AA\
region. Trager et al. (1998), and references therein, have extensively shown
the usefulness of the Lick/IDS absorption-line index system in the study of
old stellar populations.  In brief, the Lick/IDS library contains 460 stars of
all spectral types and luminosity classes. Although a large fraction are field
stars from the solar neighbourhood, members of open clusters (covering a wide
range of ages) and galactic globular clusters (with different metallicities)
are also included.

\begin{figure}
 \begin{center}
  \resizebox{\hsize}{!}{\includegraphics[angle=-90]{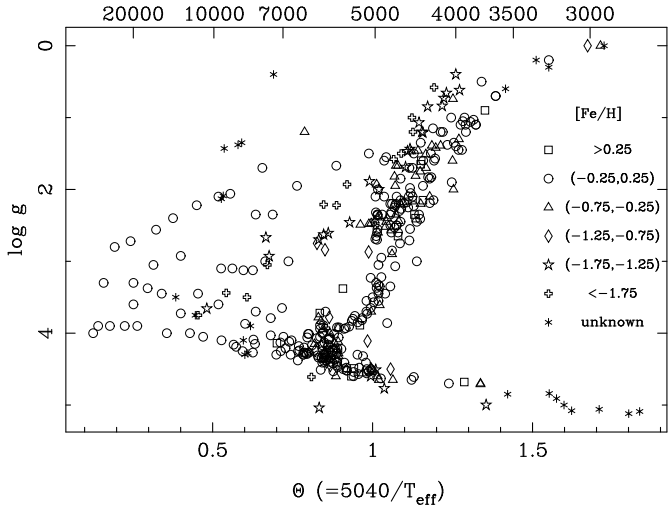}}
  \caption{Gravity--temperature diagram for the sample of stars used to 
derive the empirical fitting function. Different symbols are used to indicate 
stars of different metallicities, as shown in the key.}
  \label{hrdiag}
 \end{center}
\end{figure}

Table~\ref{tabletotal} lists the final data sample, which comprises 392 stars
out of the original set of 424 objects with published atmospheric parameters
in W94 (the remaining 32 stars presented high uncertainties in the derived
measurements and were not included in order to guarantee the quality of the
final dataset). The first two columns list the Henry Draper Catalog number, if
any, and other common designation. The spectral type and luminosity class are
given in the third column (see references at the end of the table). The
following three columns list the atmospheric parameters employed in this work.
Most of them are those listed in the electronic table of
W94. Since we wish to keep the consistency with the previous fitting functions
for the Lick indices, we have been reluctant to change any atmospheric
parameter.  However, we have introduced new parameters for some of the stars
which lack them in W94, and, in some cases, the original parameters have been
improved using more recent or reliable data from the literature (see the
description in the table notes). As a whole, effective temperature, gravity
and metallicity data are finally available for 383, 382 and 354 stars,
respectively.

To summarize, the Lick/IDS library spans the following ranges in atmospheric
parameters: $T_{\rm eff}$ from 2747 to 21860~K, $\log g$ from 0 to 5.12, and
[Fe/H] from $-2.70$ to 0.46. It is important to note that the star sample does
not homogeneously cover the parameter space, being more densely populated in
the metallicity interval $-$1.0~dex to +0.5~dex, whereas the gravity range is
wider close to solar metallicity. A more important deficiency of the library is
the paucity of hot stars. Since we are also interested in applying the derived
fitting functions to the modeling of younger stellar populations, we have
enlarged the sample with the inclusion of additional hot stars. For this
purpose, we have employed flux-calibrated spectra corresponding to
representative spectral types and luminosity classes from the compilation of
Pickles (1998). In particular, 43 spectra of O5--F0 stars 
with $6720<T_{\rm
eff}<39800$~K have been incorporated. It is worth noting that we have checked
that no systematic offset is apparent between our D4000 measurements and those
in Pickles' spectra. To display the range of stellar parameters covered by the
resulting library, in Fig~\ref{hrdiag} we present a $\log g$--$T_{\rm eff}$
diagram of the calibrating stars.

\begin{table*}
\caption{Data sample, stellar atmospheric parameters employed to derive the
fitting function, and break measurements.} 
\label{tabletotal}
{\small
\begin{center}
\begin{tabular}{rllrrrrcrl} \hline \hline
\multicolumn{1}{c}{\raisebox{0.ex}[3.0ex][1.5ex]{HD}} & Name & Type$^a$ & 
\multicolumn{1}{c}{$T_{\rm eff}$\xx{b}} & 
\multicolumn{1}{c}{$\log g$\xx{b}} & [Fe/H]\xx{b} & 
\multicolumn{1}{c}{D$_{4000}$} & ${\Delta{\rm D}_{4000}}^c$ & 
\multicolumn{1}{c}{Res.$^d$} & Runs$^e$ \\ \hline
\raisebox{0.ex}[3.0ex]{ }       
        & Coma A   3    & G9 V            &  5100        & 4.530       & 
$-0.07$       & 2.398 & 0.095 & $-0.075$ & 6               \\ 
        & Coma A  13    & K0 V            &  5357        & 4.540       & 
$-0.07$       & 2.169 & 0.094 & $-0.009$ & 6               \\
        & Coma A  14    & G4 V            &  5196        & 4.320       & 
$-0.07$       & 1.888 & 0.092 & $-0.467$ & 6               \\
        & Coma A  21    & G7 V            &  5147        & 4.410       & 
$-0.07$       & 2.137 & 0.093 & $-0.277$ & 6               \\
107214  & Coma T  65    & G0 V            &  5960        & 4.300       & 
$-0.07$       & 1.612 & 0.091 & $-0.094$ & 6               \\
107276  & Coma T  68    & A6 IV-V         &  8030        & 4.090       & 
$-0.07$       & 1.456 & 0.090 & $-0.040$ & 6               \\
107513  & Coma T  82    & A9 V            &  7080        & 4.130       & 
$-0.07$       & 1.435 & 0.064 & $-0.027$ & 6(2)            \\
\hline				         			       
\end{tabular}
\end{center}			         			       
}
\end{table*}

\addtocounter{table}{-1}
\begin{table*}
\caption{Continued.}
{\small
\begin{center}
\begin{tabular}{rllrrrrcrl} \hline \hline
\multicolumn{1}{c}{\raisebox{0.ex}[3.0ex][1.5ex]{HD}} & Name & Type$^a$ & 
\multicolumn{1}{c}{$T_{\rm eff}$\xx{b}} &
\multicolumn{1}{c}{$\log g$\xx{b}} & [Fe/H]\xx{b} & 
\multicolumn{1}{c}{D$_{4000}$} & ${\Delta{\rm D}_{4000}}^c$ & 
\multicolumn{1}{c}{Res.$^d$} & Runs$^e$ \\ \hline
\raisebox{0.ex}[3.0ex]{ }       
        & M67 F 084     & {\it HB }       &  4763        & 2.320       & 
$-0.10$       & 2.936 & 0.098 & $ 0.081$ & 6               \\
        & M67 F 094     & {\it SGB}       &  6103        & 4.070       & 
$-0.10$       & 1.698 & 0.092 & $ 0.078$ & 6               \\
        & M67 F 105     & {\it GB }       &  4465        & 2.230       & 
$-0.10$       & 3.411 & 0.102 & $ 0.158$ & 6               \\
        & M67 F 108     & {\it GB }       &  4258        & 1.830       & 
$-0.10$       & 3.782 & 0.103 & $ 0.216$ & 6               \\
        & M67 F 115     & {\it SGB}       &  5959        & 3.890       & 
$-0.10$       & 1.763 & 0.093 & $ 0.066$ & 6               \\
        & M67 F 117     & {\it SGB}       &  5210        & 3.790       & 
$-0.10$       & 1.898 & 0.066 & $-0.421$ & 6(2)            \\
        & M67 F 119     & {\it SGB}       &  6031        & 3.940       & 
$-0.10$       & 1.683 & 0.092 & $ 0.026$ & 6               \\
\hline				         			       
\end{tabular}
\end{center}			         			       
}
\end{table*}

\addtocounter{table}{-1}
\begin{table*}
\caption{Continued.}
{\small
\begin{center}
\begin{tabular}{rllrrrrcrl} \hline \hline
\multicolumn{1}{c}{\raisebox{0.ex}[3.0ex][1.5ex]{HD}} & Name & Type$^a$ & 
\multicolumn{1}{c}{$T_{\rm eff}$\xx{b}} &
\multicolumn{1}{c}{$\log g$\xx{b}} & [Fe/H]\xx{b} & 
\multicolumn{1}{c}{D$_{4000}$} & ${\Delta{\rm D}_{4000}}^c$ & 
\multicolumn{1}{c}{Res.$^d$} & Runs$^e$ \\ \hline
\raisebox{0.ex}[3.0ex]{ }       
        & M92 XII-8     & {\it GB }       &  4488        & 1.000       & 
$-2.20$       & 1.843 & 0.069 & $-0.316$ & 6(2)            \\
        & M92 XII-24    & {\it HB }       & 11100        & 3.750       & 
$-2.20$       & 1.498 & 0.060 & $ 0.235$ & 9,12            \\
        & NGC188 I-20   & {\it    }       &              &             & 
$     $       & 2.605 & 0.060 & \multicolumn{1}{c}{$\dag$} & 10              \\
        & NGC188 I-55   & {\it SGB}       &  5440        & 3.910       & 
$ 0.00$       & 2.076 & 0.043 & $-0.063$ & 9,10            \\
        & NGC188 I-57   & {\it GB }       &  4707        & 3.070       & 
$ 0.00$       & 3.035 & 0.047 & $ 0.029$ & 9,10            \\
        & NGC188 I-61   & {\it GB }       &  4930        & 3.350       & 
$ 0.00$       & 2.984 & 0.061 & $ 0.245$ & 10              \\
        & NGC188 I-69   & {\it GB }       &  4401        & 2.350       & 
$ 0.00$       & 3.479 & 0.044 & $ 0.066$ & 9,10            \\
\hline				         			       
\end{tabular}
\end{center}			         			       
}
\end{table*}

\addtocounter{table}{-1}
\begin{table*}
\caption{Continued.}
{\small
\begin{center}
\begin{tabular}{rllrrrrcrl} \hline \hline
\multicolumn{1}{c}{\raisebox{0.ex}[3.0ex][1.5ex]{HD}} & Name & Type$^a$ & 
\multicolumn{1}{c}{$T_{\rm eff}$\xx{b}} &
\multicolumn{1}{c}{$\log g$\xx{b}} & [Fe/H]\xx{b} & 
\multicolumn{1}{c}{D$_{4000}$} & ${\Delta{\rm D}_{4000}}^c$ & 
\multicolumn{1}{c}{Res.$^d$} & Runs$^e$ \\ \hline
\raisebox{0.ex}[3.0ex]{7927}       
        & HR 0382       & F0 Ia           &  7300        & 0.400       & 
$     $       & 1.523 & 0.090 & $ 0.053$ & 6               \\
  8424  & HR 0398       & A0 Vnn          &  8455        & 4.100       & 
$     $       & 1.400 & 0.090 & $-0.054$ & 6               \\
 10307  & HR 0483       & G2 V            &  5898        & 4.380\xx{4} & 
$ 0.04$\xx{4} & 2.132 & 0.086 & $ 0.351$ & 5               \\
 10380  & HR 0489       & K3 III          &  4133        & 1.200       & 
$-0.11$       & 4.323 & 0.095 & $ 0.561$ & 3               \\
 10476  & HR 0493       & K1 V            &  5065        & 4.500\xx{1} & 
$-0.20$\xx{5} & 2.387 & 0.137 & $-0.053$ & 4               \\
 10700  & HR 0509       & G8 V            &  5200        & 4.560       & 
$-0.37$       & 1.892 & 0.086 & $-0.276$ & 5               \\
 10780  & HR 0511       & K0 V            &  5398        & 4.600\xx{1} & 
$ 0.36$\xx{1} & 2.319 & 0.136 & $-0.109$ & 4               \\
\hline				         			       
\end{tabular}
\end{center}			         			       
}
\end{table*}

\addtocounter{table}{-1}
\begin{table*}
\caption{Continued.}
{\small
\begin{center}
\begin{tabular}{rllrrrrcrl} \hline \hline
\multicolumn{1}{c}{\raisebox{0.ex}[3.0ex][1.5ex]{HD}} & Name & Type$^a$ & 
\multicolumn{1}{c}{$T_{\rm eff}$\xx{b}} &
\multicolumn{1}{c}{$\log g$\xx{b}} & [Fe/H]\xx{b} & 
\multicolumn{1}{c}{D$_{4000}$} & ${\Delta{\rm D}_{4000}}^c$ & 
\multicolumn{1}{c}{Res.$^d$} & Runs$^e$ \\ \hline
\raisebox{0.ex}[3.0ex]{48682}       
        & HR 2483       & G0 V            &  5829        & 4.080       & 
$ 0.15$\xx{1} & 1.717 & 0.059 & $-0.158$ & 6,11(2)         \\
 49161  & HR 2503       & K4 III          &  4276        & 1.850       & 
$ 0.23$       & 3.535 & 0.100 & $-0.206$ & 6               \\
 49293  & HR 2506       & K0 III          &  4635        & 1.850       & 
$-0.07$       & 2.990 & 0.097 & $-0.048$ & 6               \\
 50778  & HR 2574       & K4 III          &  4019        & 1.350       & 
$-0.13$       & 4.413 & 0.107 & $ 0.466$ & 6               \\
 51440  & HR 2600       & K2 III          &  4381        & 2.200       & 
$-0.35$       & 3.227 & 0.061 & $ 0.007$ & 6,11(2),13      \\
 54719  & HR 2697       & K2 III          &  4381        & 1.350       & 
$ 0.24$       & 3.465 & 0.099 & $-0.142$ & 6               \\
 54810  & HR 2701       & K0 III          &  4697        & 2.350       & 
$-0.30$       & 2.650 & 0.095 & $-0.157$ & 6               \\
\hline				         			       
\end{tabular}
\end{center}			         			       
}
\end{table*}

\addtocounter{table}{-1}
\begin{table*}
\caption{Continued.}
{\small
\begin{center}
\begin{tabular}{rllrrrrcrl} \hline \hline
\multicolumn{1}{c}{\raisebox{0.ex}[3.0ex][1.5ex]{HD}} & Name & Type$^a$ & 
\multicolumn{1}{c}{$T_{\rm eff}$\xx{b}} &
\multicolumn{1}{c}{$\log g$\xx{b}} & [Fe/H]\xx{b} & 
\multicolumn{1}{c}{D$_{4000}$} & ${\Delta{\rm D}_{4000}}^c$ & 
\multicolumn{1}{c}{Res.$^d$} & Runs$^e$ \\ \hline
\raisebox{0.ex}[3.0ex]{114762}       
        &               & F9 V            &  5814        & 4.260       & 
$-0.75$\xx{4} & 1.379 & 0.042 & $-0.190$ & 3               \\
114946  & HR 4995       & G6 V            &  4930        & 3.610       & 
$-0.44$       & 2.768 & 0.096 & $ 0.239$ & 6               \\
114961  &               & M7 III          &  3014        & 0.000       & 
$-0.81$       & 2.633 & 0.095 & $-0.046$ & 6               \\
115043  &               & G1 V            &  6016        & 4.510       & 
$-0.08$\xx{1} & 1.705 & 0.046 & $ 0.034$ & 3               \\
115617  & HR 5019       & G6 V            &  5480        & 4.520       & 
$ 0.22$       & 1.976 & 0.050 & $-0.262$ & 3               \\
117176  & HR 5072       & G5 V            &  5370        & 3.800       & 
$-0.21$       & 1.990 & 0.033 & $-0.091$ & 3,6(2),8(2)     \\
120136  & HR 5185       & F7 V            &  6431        & 4.270       & 
$ 0.18$\xx{1} & 1.497 & 0.043 & $-0.027$ & 3               \\
\hline				         			       
\end{tabular}
\end{center}			         			       
}
\end{table*}

\addtocounter{table}{-1}
\begin{table*}
\caption{Continued.}
{\small
\begin{center}
\begin{tabular}{rllrrrrcrl} \hline \hline
\multicolumn{1}{c}{\raisebox{0.ex}[3.0ex][1.5ex]{HD}} & Name & Type$^a$ & 
\multicolumn{1}{c}{$T_{\rm eff}$\xx{b}} &
\multicolumn{1}{c}{$\log g$\xx{b}} & [Fe/H]\xx{b} & 
\multicolumn{1}{c}{D$_{4000}$} & ${\Delta{\rm D}_{4000}}^c$ & 
\multicolumn{1}{c}{Res.$^d$} & Runs$^e$ \\ \hline
\raisebox{0.ex}[3.0ex]{157214}
        & HR 6458       & G1 V            &  5686        & 4.240\xx{6} & 
$-0.41$\xx{6} & 1.655 & 0.041 & $-0.088$ & 3,6             \\
160693  &               & G0 V            &  5733        & 4.220       & 
$-0.67$\xx{5} & 1.632 & 0.085 & $ 0.007$ & 5               \\
161797  & HR 6623       & G5 IV           &  5390        & 3.830       & 
$ 0.23$       & 2.289 & 0.093 & $-0.053$ & 6               \\
161817  &               & A2 VI           &  7450        & 2.930       & 
$-1.71$       & 1.377 & 0.090 & $-0.103$ & 6               \\
163506  & HR 6685       & F2 Ibe          &  6400        & 1.200       & 
$-0.41$       & 1.425 & 0.064 & $-0.164$ & 6(2)            \\
164259  & HR 6710       & F3 V            &  6750        & 4.100       & 
$-0.12$\xx{1} & 1.428 & 0.061 & $-0.017$ & 5,6             \\
165195  &               & K3p             &  4475        & 1.200       & 
$-2.70$       & 2.122 & 0.083 & $ 0.115$ & 2,5             \\
\hline				         			       
\end{tabular}
\end{center}			         			       
}
\end{table*}

\addtocounter{table}{-1}
\begin{table*}
\caption{Continued.}
{\small
\begin{center}
\begin{tabular}{rllrrrrcrl} \hline \hline
\multicolumn{1}{c}{\raisebox{0.ex}[3.0ex][1.5ex]{HD}} & Name & Type$^a$ & 
\multicolumn{1}{c}{$T_{\rm eff}$\xx{b}} &
\multicolumn{1}{c}{$\log g$\xx{b}} & [Fe/H]\xx{b} & 
\multicolumn{1}{c}{D$_{4000}$} & ${\Delta{\rm D}_{4000}}^c$ & 
\multicolumn{1}{c}{Res.$^d$} & Runs$^e$ \\ \hline
\raisebox{0.ex}[3.0ex]{217476}       
        & HR 8752       & G4 Ia           &  5100        & 1.500\xx{7} & 
$ 0.07$\xx{1} & 1.635 & 0.091 & $-0.949$ & 6               \\
219734  & HR 8860       & M2 III          &  3730        & 0.900       & 
$ 0.27$\xx{14}& 3.990 & 0.104 & $-0.003$ & 6               \\
222368  & HR 8969       & F7 V            &  5959        & 3.940       & 
$-0.25$\xx{6} & 1.644 & 0.133 & $-0.009$ & 4               \\
232078  &               & K4-5 III-II     &  4000        & 0.400       & 
$-1.60$       & 3.989 & 0.100 & $ 0.384$ & 6,11(2)         \\
232979  & Gl 172        & K8 V            &  3769        & 4.700       & 
$-0.33$\xx{14}& 2.959 & 0.091 & $ 0.146$ & 5               \\
        & BD+ 17 4708   & F8 VI           &  5900        & 3.930       & 
$-1.90$       & 1.368 & 0.078 & $ 0.064$ & 8               \\
        & BD+ 19 5116A  & M4 V            &  3200        & 4.910       & 
$     $       & 2.310 & 0.070 & $ 0.158$ & 6(2)            \\
\hline
\end{tabular}
\vspace{1.5mm}

\parbox{15.4cm}{
$^a$ Sources for spectral types are, in order of preference: W94, G93, the 
Bright Star Catalog (Hoffleit 1982) and the Simbad database. For the cluster
stars (except for Coma and Hyades) we list positions in the HR diagram ({\it
SGB\/}: subgiant branch; {\it GB\/}: giant branch; {\it HB\/}: horizontal
branch, red clump stars; {\it AGB\/}: asymptotic giant branch). 

$^b$ Atmospheric parameters from the electronic table of W94, with the
following additional sources and modifications: 
 (1) Cayrel de Strobel et al. (1997). (2) Alonso et al. (1996). (3) Gratton et
 al. (1996). (4) Mean from sources 1, 2 and 3. (5) Mean from sources 1 and
 2. (6) Mean from sources 1 and 3. (7) Thevenin (1998). (8) Marsakov and
 Shevelev (1995). (9) Gies and Lambert (1992). (10) Zakhozhaj and Shaparenko
 (1996). (11) Dyck et al. (1996). (12) Computed from the indices Fe5270
 and Fe5335 using the fitting functions of G93 (in terms of $(V-K)$). (13)
 Computed from the indices Fe5270 and Fe5335 using the fitting functions of
 W94 (in terms of $T_{\rm eff}$). (14) Computed from the index Fe5270 using the
 fitting functions of W94 (in terms of $T_{\rm eff}$). (15) From spectral type
 using table III of Johnson (1966). (16) Extrapolation in the relation
 $T_{\rm eff}$ versus spectral type of Ridgway et al (1980) ---these 
 temperatures differ from those given by W94---.

$^c$ Total random errors in the $\lambda4000$~\AA\ break measurements.

$^d$ Residuals from the derived fitting functions (observed $-$
predicted). Stars not used to derive the empirical functions are marked with
$\dag$ (non cluster members, see G93), $\dag\dag$ (lacking required
atmospheric parameters) or $\dag\dag\dag$ (emission-line supergiants).

$^e$ Run number(s) in which the star was observed. Number of repeated
observations within each run is giving in parenthesis.
}

\end{center}
}
\end{table*}

\section{Observations and data reduction}

The stellar library was observed in a total of 
13 observing runs from 1991 to 1996 using the JKT, INT and WHT at the Roque de 
los Muchachos Observatory (La Palma, Spain), the
3.5~m telescope at Calar Alto Observatory (Almer\'{\i}a, Spain), and the 
2.12~m telescope at San Pedro M\'{a}rtir Observatory (Mexico). The bulk of the 
spectra (320 out of a total of 650 spectra) were taken in run~6, where
the Richardson--Brealey Spectrograph was used to obtain 2.8 \AA\ resolution 
spectra in the blue spectral range.    
A description of each observing run, including relevant observational
configuration parameters, is given in Table~\ref{tableruns}. The number of
stars observed in each run is quite variable since some of the runs were not
devoted to the calibration of the break and, in these cases, only bright stars
of the library were observed during twilight periods. The last column in
Table~\ref{tabletotal} list the run numbers in which each star was observed.

\begin{table*}
\caption{Observational configurations.}
\label{tableruns}
\begin{center}
\begin{tabular}{cccccccc} \hline\hline
\raisebox{0.ex}[3.0ex][1.5ex]{Run} & Dates & Telescope$^a$ & Spectrograph & 
Detector & Disp.$^b$ & $\Delta\lambda\ ^c$ & FWHM$^d$ \\
\hline
\raisebox{0.ex}[3.0ex]{1}       
       & 1--3 Nov 1991   & WHT 4.2m   & ISIS blue arm & EEV\#6   & 2.70     &
 3500--6850       &   16.0              \\
 2     & 8--10 Jun 1993  & CAHA 3.5m  & CTS           & TEK\#12  & 3.46     &
 3750--7350       &    7.2              \\
 3     & 2--6 Jul 1994   & INT 2.5m   & IDS 235mm     & TEK\#3   & 3.30     &
 3700--7100       &    5.2              \\
 4     & 9--10 Aug 1994  & CAHA 3.5m  & CTS           & TEK\#12  & 3.46     &
 3700--7240       &    8.1              \\
 5     & 10--14 Sep 1994 & INT 2.5m   & IDS 235mm     & TEK\#3   & 3.30     &
 3700--7100       &    5.8              \\
 6     & 13--19 Feb 1995 & JKT 1.5m   & RBS           & TEK\#4   & 0.91     &
 3600--4530       &    2.8              \\
 7     & 15 Feb 1995     & INT 2.5m   & IDS 235mm     & TEK\#3   & 0.85     &
 3634--4502       &    2.1              \\
 8     & 17--19 Dec 1995 & WHT 4.2m   & ISIS blue arm & TEK\#1   & 2.90     &
 3740--6700       &   12.3              \\
 9     & 15--16 Jan 1996 & CAHA 3.5m  & CTS           & TEK\#11  & 0.88     &
 3654--4554       &    2.5              \\
10     & 4 Aug 1996      & WHT 4.2m   & ISIS blue arm & TEK\#2   & 1.53     &
 3654--5219       &    4.9              \\
11     & 28 Mar--1 Apr 1995 & SPM 2.12m& B\&Ch   & TEK      & 4.00     &
 3500--7592       &    8.3              \\
12     & 3 Mar 1995      & INT 2.5m   & IDS 500mm     & TEK\#3   & 0.75     &
 3670--4440       &    4.2              \\
13     & 17--18 Nov 1996 & CAHA 3.5m  & CTS           & SITe\#6a & 1.10     &
 3569--5765       & \raisebox{0.ex}[0.ex][1.5ex]{3.4} \\
\hline
\multicolumn{8}{l}{$^a$ JKT, INT and WHT (La Palma, Spain),
\raisebox{0.ex}[3.0ex]{CAHA} (Calar Alto, Spain), 
SPM (San Pedro M\'{a}rtir, Mexico)} \\
\multicolumn{8}{l}{$^b$ dispersion (\AA/pixel)} \\
\multicolumn{8}{l}{$^c$ wavelength range (\AA)} \\
\multicolumn{8}{l}{$^d$ measured spectral resolution (\AA)} \\
\end{tabular}
\end{center}
\end{table*}

%
%
%

The reduction of the data was performed with our own reduction package
\reduceme\footnote{See description of this package in:\\
http://www.ucm.es/info/Astrof/reduceme/reduceme.html} (Cardiel \& 
Gorgas 1999), which allows a parallel 
treatment of data and error spectra (see below). We followed a standard
reduction procedure for spectroscopic data: bias and dark subtraction, cosmic
ray cleaning, flat-fielding, wavelength calibration, C-distortion correction,
sky subtraction, atmospheric extinction correction and flux calibration (we
did not attempt to obtain absolute fluxes since, as most line-strength
indices, the break only requires relative fluxes). Cluster stars spectra were
also corrected from interstellar reddening, using the color excesses quoted in
Tables~4 and~A3 from G93 and W94, respectively, and the averaged extinction
curve of Savage \& Mathis (1979).

In order to optimize the observing time during most of the runs, comparison
arc frames were not taken next to each star observation. Instead of this, we
only acquired arc exposures for a selected subsample of stars, which comprised
objects with a complete coverage of all the spectral types and luminosity
classes observed in each run. The wavelength calibration of the rest of the
stars was performed by comparing them with the reference spectra.  The
repetition of this procedure with different reference spectra allow us to
guarantee that wavelength calibration errors were always $\la 0.2$~pixels.

\section{Random errors and systematic effects}
\label{secerr}

Since the aim of this paper is to derive an analytical representation of the
behaviour of the D$_{4000}$ as a function of effective temperature,
metallicity and surface gravity, the sources of error are two-fold. In one
hand, an important error source are the uncertainties in the adopted
atmospheric stellar parameters. Detailed discussions of the sources of the
stellar parameters and their associated errors are given in the original
papers G93 and W94. In this work we assume that these errors are random and,
thus, their effect in the fitting procedure is minimized through the use of a
library containing a large number of stars. The other type of errors are those
associated to D$_{4000}$ measurements, which are the subject of this
section. Undoubtly, an accurate knowledge of the errors is essential to
guarantee the validity of the final product of this work, i.e., the fitting
functions of the break.

Since, apart from the cluster members, most of the stars of the Lick/IDS
library are bright, and considering the low signal-to-noise ratio required to
measure the D$_{4000}$ with acceptable accuracy, systematic errors are the
main source of uncertainty. 

\subsection{Random errors}

{\sl (i) Photon statistics and read-out noise.} With the aim of tracing the
propagation of photon statistics and read-out noise, we followed a parallel
reduction of data and error frames. For a detailed description on the
estimation of random errors in the measurement of line-strength indices we
refer the interested reader to Cardiel et al. (1998). Starting with the
analysis of the photon statistics and read-out noise, the reduction package
\reduceme\ is able to generate error frames from the beginning of the reduction
procedure, and properly propagates the errors throughout the reduction process.
In this way, important reduction steps such as flatfielding, geometrical
distortion corrections, wavelength calibration and sky subtraction, are taken
into account. At the end of the reduction process, each data spectrum
$S(\lambda_i)$ has its associated error spectrum $\sigma(\lambda_i)$, which can
be employed to derive accurate index errors. 
The errors in the break are computed by (Cardiel et al. 1998)
\begin{equation}
\Delta^2[{\rm D}_{4000}]_{\rm photon} =
  \frac{ {\cal F}_r \sigma^2_{{\cal F}_b}+ {\cal F}_b \sigma^2_{{\cal F}_r} }%
       { {\cal F}_b^4 },
\label{eqerrd4000a}
\end{equation}
with
\begin{equation}
{\cal F}_p \equiv \sum_{i=1}^{N_p} [\lambda_i^2 S(\lambda_i)],
\label{eqerrd4000b}
\end{equation}
and
\begin{equation}
\sigma^2_{{\cal F}_p} = 
   \Theta^2 \sum_{i=1}^{N_p} [\lambda_i^4 \sigma^2(\lambda_i)],
\label{eqerrd4000c}
\end{equation}
where the subscripts $b$ and $r$ correspond, respectively, to the blue and red
bandpasses of the break ($p$ refers indistinctly to $b$
or $r$), $S(\lambda_i)$ and $\sigma(\lambda_i)$ are the signal and the error in
the pixel with central wavelength $\lambda_i$, $\Theta$ is the
dispersion (in \AA/pixel) assuming a linear wavelength scale, and $N_p$ is the
number of pixels covered by the $p$ band (in general, fractions of pixels must
be considered at the borders of the bandpasses). We have checked that the 
above analytical formulae exhibit an excellent agreement with numerical 
simulations. For the whole sample, the error of a typical observation
introduced by these sources of noise is 
$\langle\Delta\left[{\rm D}_{4000}\right]_{\rm photon}\rangle=0.038$.

{\sl (ii) Flux calibration.} During each run we observed a number (typically
around 5) of different
spectrophotometric standard stars (from Massey et al. 1988 and Oke 1990). 
The break was measured using the average flux calibration curve, and
we estimated the random error in flux calibration as the r.m.s. scatter among
the different D$_{4000}$ values obtained with each standard. The typical error
introduced by this uncertainty is $\langle\Delta\left[{\rm
D}_{4000}\right]_{\rm flux}\rangle=0.034$.

{\sl (iii) Wavelength calibration and radial velocity correction.} These two
reduction steps are potential sources of random errors in the wavelength scale
of the reduced spectra. Radial velocities for field stars were obtained from
the Hipparcos Input Catalogue (Turon et al. 1992), which in the worst cases
are given with mean probable errors of $\sim 5$~km~s$^{-1}$ ($\sim 0.07$~\AA\
at $\lambda$4000~\AA). For the cluster stars, we used either published radial
velocities for individual stars, if available, or averaged cluster radial
velocities (Hesser et al. 1986: M3, M5, M10, M13, M71, M92, NGC6171; Friel
1989: NGC188; Friel \& Janes 1993: M67, NGC7789; Turon et al. 1992: Coma,
Hyades). Typical radial velocity errors for the cluster stars are $\la
15$~km~s$^{-1}$ ($\sim 0.2$~\AA\ at $\lambda$4000~\AA).  To have an estimate
of the random error introduced by the combined effect of wavelength
calibration and radial velocity, we cross-correlated fully calibrated spectra
corresponding to stars of similar spectral types. The resulting typical error
is 20~km~s$^{-1}$, being always below 75~km~s$^{-1}$. This translates into a
negligible error of $\langle\Delta\left[{\rm D}_{4000}\right]_{\rm
wavelength}\rangle=0.003$. 
However, it may be useful to estimate the importance of this
effect when measuring the break in galaxies with large radial velocity
uncertainties. As a reference, using the 18 spectra displayed in
Fig.~\ref{plotspt}, a velocity shift of $\sim 100$~km~s$^{-1}$ translates into
relative D$_{4000}$ errors always below 1\%. Furthermore, for K0~III stars we
obtain
\mbox{$\Delta[{\rm D}_{4000}]_{\rm wavelength} \simeq 1.56 \times 10^{-4} 
\Delta v$}, where $\Delta v$ is the velocity error in km~s$^{-1}$ (this
relation only holds for $\Delta v \le 150$~km~s$^{-1}$; for $\Delta v$
in the range from 150--1000 km~s$^{-1}$ the error increases slower, and remains
below $0.1$).

{\sl (iv) Additional sources of random errors.} Expected random errors for each
star can be computed by adding quadratically the random errors derived from
the three sources previously discussed, i.e.,

\begin{displaymath}
\Delta^2[{\rm D}_{4000}]_{\rm expected}=
\end{displaymath}
\begin{equation}
  \Delta^2[{\rm D}_{4000}]_{\rm photon}+
  \Delta^2[{\rm D}_{4000}]_{\rm flux}+
  \Delta^2[{\rm D}_{4000}]_{\rm wavelength}.
\end{equation}

However, additional (and unknown) sources of random error may 
still be present in the data. 
Following the method described in Gonz\'{a}lez (1993), we compared, within each
run, the standard deviation of the D$_{4000}$ measurements of stars with
multiple observations with the expected error 
$\Delta[{\rm D}_{4000}]_{\rm expected}$. 
For those runs in which the standard deviation was significantly
larger than the expected error (using the $F$-test of variances with a
significance level $\alpha=0.3$), a residual random error $\Delta_{\rm
residual}$ was derived and added to all the individual stellar random
\mbox{errors:}
\pagebreak
\begin{displaymath}
\Delta^2[{\rm D}_{4000}]_{\rm random} =
\end{displaymath}
\begin{equation}
\Delta^2[{\rm D}_{4000}]_{\rm expected}+
\Delta^2[{\rm D}_{4000}]_{\rm residual}.
\label{err_random_uno}
\end{equation}
 
It is worth noting that this additional error was only needed for some
runs. In the particular case of run~6, with a large number (54)
of stars with multiple observations, the agreement between expected and
measured error was perfect.

\subsection{Systematic effects}

The main sources of systematic effects in the measurement of spectral indices
in stars are spectral resolution, sky subtraction and flux calibration.

{\sl (i) Spectral resolution.} We have examined the effect of instrumental
broadening in the break by convolving the 18 spectra of Fig.~\ref{plotspt}
with a broadening function of variable width. The result of this study
indicates that, as expected, the break is quite insensitive to spectral
resolution.  As a reference, for a spectral resolution of 30\AA\ (FWHM) the
effect in the break is below 1\%. Therefore, given the resolutions used in
this work (last column in Table~\ref{tableruns}) no corrections are needed in
any case.

{\sl (ii) Sky subtraction.} Since the field giant and dwarf stars of the
library are bright, the exposure times were short enough to neglect the effect
of an anomalous subtraction of the sky level. However, most of the cluster
stars are not bright, being necessary exposures times of up to 1800 seconds
for the faintest objects. In addition, the observation of these stars,
specially those in globular clusters, were performed with the unavoidable
presence of several stars inside the spectrograph slit, which complicated the
determination of the sky regions. In Cardiel et al. (1995) we already studied
the systematic variations on the D$_{4000}$ measured in the outer parts of a
galaxy (where light levels are only a few per cent of the sky signal) due to
the over-- or under--estimation of the sky level. 
We refer the interested reader to that paper for details. Although
there is not a simple recipe to detect this type of systematic effect,
unexpectedly high D$_{4000}$ values in faint cluster stars can arise from an
anomalous sky subtraction.

{\sl (iii) Flux calibration.} Due to the large number of runs needed to
complete the whole library, important systematic errors can arise due to
possible differences among the spectrophotometric system of each run. In order
to guarantee that the whole dataset is in the same system, we compared the
measurements of the stars in common among different runs.  Since run~6 was the
observing run with the largest number of stars (including numerous multiple
observations) and with reliable random errors (see above), we selected it as
our spectrophotometric reference system. Therefore, for each run we computed a
mean offset with run 6, which was introduced when it was significantly
different from 0 (using a $t$ test).  It is important to highlight that
differences between a {\it true\/} spectrophotometric system and that
adopted in this work may still be present.  Therefore, we encourage the readers
interested in the predictions of the present fitting functions, to include in
their observations a number of template stars from the library to ensure a
proper correction of the data.

\subsection{Final errors}

The comparison of measurements of the same stars in different runs also
provides a powerful method to refine the random errors derived in
Eq.~\ref{err_random_uno}.  We followed an iterative method which consistently
provided the relative offsets and a set of extra residual errors to account for
the observed scatter among runs (see Cardiel 1999 for details).

As mentioned before, the data sample was enlarged by including 43~stellar
spectra from the Pickles' (1998) library. The random errors in the D$_{4000}$
indices measured in this subsample were estimated from the residual variance of
a least-square fit to a straight line using all the stars (except supergiants)
with $T_{\rm eff}>8400$ K (they follow a tight linear relation in the
D$_{4000}$--$\theta$ plane). The typical error in Pickles' spectra was found to
be 0.036.    

\section{D$_{4000}$ measurements}

\begin{figure*}
 \begin{center} 
  \resizebox{130mm}{!}{\includegraphics{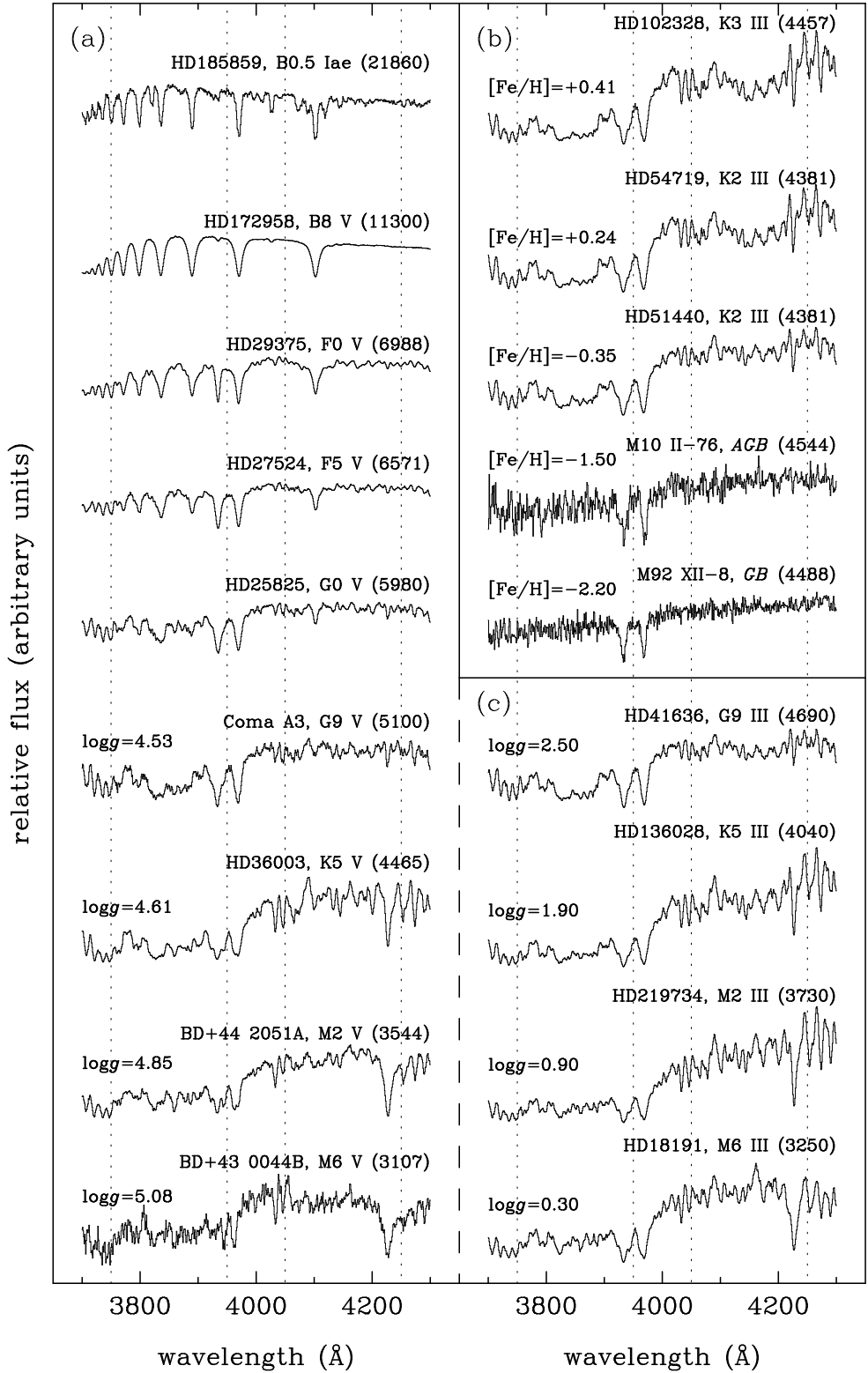}}
  \caption{Sample spectra of stars observed in run~6. Effective temperatures
 are given in parenthesis. Panel~(a) is a sequence in spectral types for main
 sequence stars. Panel~(b) shows stars with similar temperature but with a
 wide range in metallicity. Panel~(c) displays a sequence in spectral types
 for giant stars, which can be compared with the lower part of the dwarf
 sequence in panel~(a).}  
  \label{plotspt} 
 \end{center}
\end{figure*}

\begin{figure*}
 \begin{center}
  \resizebox{180mm}{!}{\includegraphics[angle=-90]{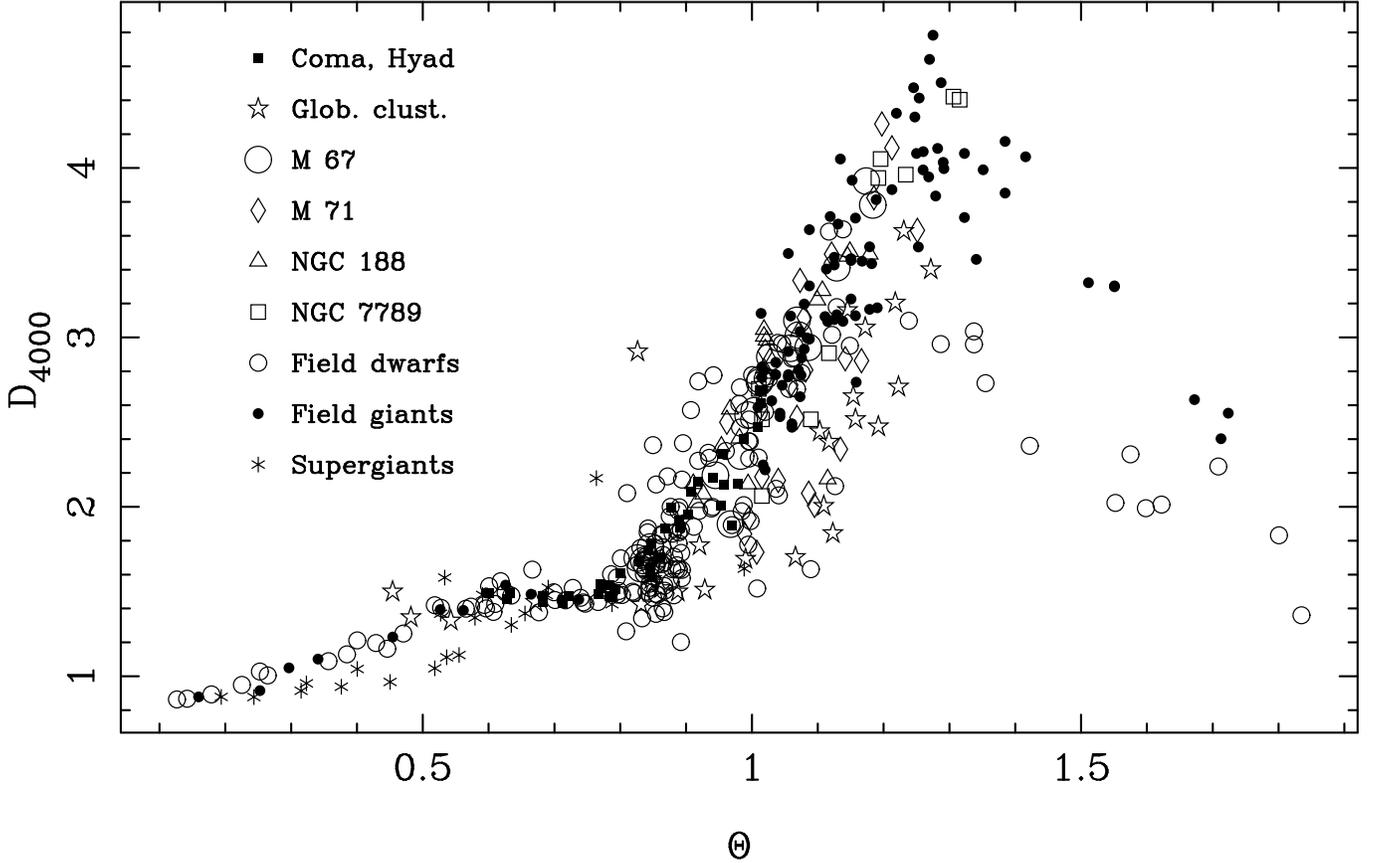}}
  \caption{D$_{4000}$ as a function of $\theta\equiv 5040/T_{\rm eff}$ for the 
whole sample. Stars are plotted using the same code as in G93 or W94}
  \label{licksymbol}
 \end{center}
\end{figure*}

Table~\ref{tabletotal} (columns 7 and 8) lists the final D$_{4000}$
measurements and the associated random errors.
Some sample spectra, exhibiting a diversity in spectral types,
metallicities and gravities, are displayed in Fig.~\ref{plotspt}, whereas in 
Fig.~\ref{licksymbol} we show the break behaviour with effective temperature
for the whole sample. It is clear from these plots that temperature is the
main parameter governing D$_{4000}$. The effect of metallicity is clearly
noticeable in panel~\ref{plotspt}(b) and by the position of globular cluster
stars in Fig.~\ref{licksymbol}. Also, some gravity dependences are also
observed, especially between hot dwarf--giants and supergiants, and 
between cold dwarfs and giants.

The definition of the D$_{4000}$ given in
Eq.~\ref{definitionD4000} resembles that of a color. However, the peculiar
combination of $\nu$ and $\lambda$ translates into the introduction of a
wavelength weighting of the flux (Eq.~\ref{definitionD4000bis}).
In order to facilitate the computation of the break, we have studied the effect
of a redefinition of the index, namely
\begin{equation}
{\rm B}_{4000} \equiv
   \frac{{\displaystyle \int_{4050\,(1+z)}^{4250\,(1+z)}} 
                        f_\lambda \; {\rm d}\lambda}%
        {{\displaystyle \int_{3750\,(1+z)}^{3950\,(1+z)}} 
                        f_\lambda \; {\rm d}\lambda}.
\label{definitionB4000}
\end{equation}

Comparing this and the previous definition of D$_{4000}$
(Eq.~\ref{definitionD4000bis}) for all the stars in run~6, we find that no
single star deviates more than a 1\% from the theoretical predicted ratio 

\begin{equation}
\left. \frac{{\rm D}_{4000}}{{\rm B}_{4000}} 
\right|_{f_{\lambda}={\rm cte}} = 1.1619,
\end{equation}
obtained for a constant $f_\lambda$. Therefore the above ratio can be safely
used to convert between both break definitions.


\section{The fitting functions}

The main aim of this work is to derive empirical fitting functions for the
D$_{4000}$ in terms of the stellar atmospheric parameters: effective
temperature, metallicity and surface gravity. After some experimentation, we
decided to use $\theta\equiv5040/T_{\rm eff}$ as the temperature indicator,
being [Fe/H] and $\log g$ the parameters for the metallicity and
gravity. Following the previous works of G93 and W94, the fitting functions are
expressed as polynomials in the atmospheric parameters, using
two different functional forms:

\begin{equation}
{\rm D}_{4000}(\theta,{\rm [Fe/H]},\log g) = 
    p(\theta,{\rm [Fe/H]},\log g)
\label{fittingforma}
\end{equation}
and
\begin{equation}
{\rm D}_{4000}(\theta,{\rm [Fe/H]},\log g) = {\rm const.} +
    e^{p(\theta,{\rm [Fe/H]},\log g)} ,
\label{fittingformb}
\end{equation}
where $p$ is a polynomial with terms up to the third order, including all
possible cross-terms among the parameters: 

\begin{equation}
p(\theta,{\rm [Fe/H]},\log g) = \sum_{k=0}^{19} c_k\ \theta^i\ {\rm [Fe/H]}^j\ 
(\log g)^l,
\label{polin}
\end{equation}
with $0\leq i+j+l\leq3$.

The polynomial coefficients were determined from a least squares fit where all
the stars were weighted according to the D$_{4000}$ observational errors
listed in Table~1. Note that this is an improvement over the 
procedure employed by G93 and W94 for the Lick indices.

Obviously, not all possible terms are necessary. The strategy followed to
determine the final fitting function is the successive inclusion of
terms, starting with the lower powers. At each step, the term which yielded a
lower new residual variance was tested. The significance of this
new term, as well as those of all the previously included coefficients, was
computed using a t-test (i.e. from the error in the coefficient, we tested
whether it was significantly different from zero). Note that this is
equivalent to performing a F-test to check whether the unbiased residual
variance is significantly reduced with the inclusion of the additional term.  
Following this procedure, and using typically a
significant level of $\alpha=0.10$, only statistically significant terms were
retained. The problem is well constrained and, usually, after the inclusion of
a few terms, a final residual variance is asymptotically reached
and the higher order terms are not statistically significant. Throughout
this fitting procedure we also kept an eye on the residuals to assure that no
systematic behaviour for any group of stars (specially stars from any given
cluster or metallicity range) was apparent. 

\begin{figure*}
 \begin{center}
  \resizebox{180mm}{!}{\includegraphics[bb= 35 416 540 737,angle=0]{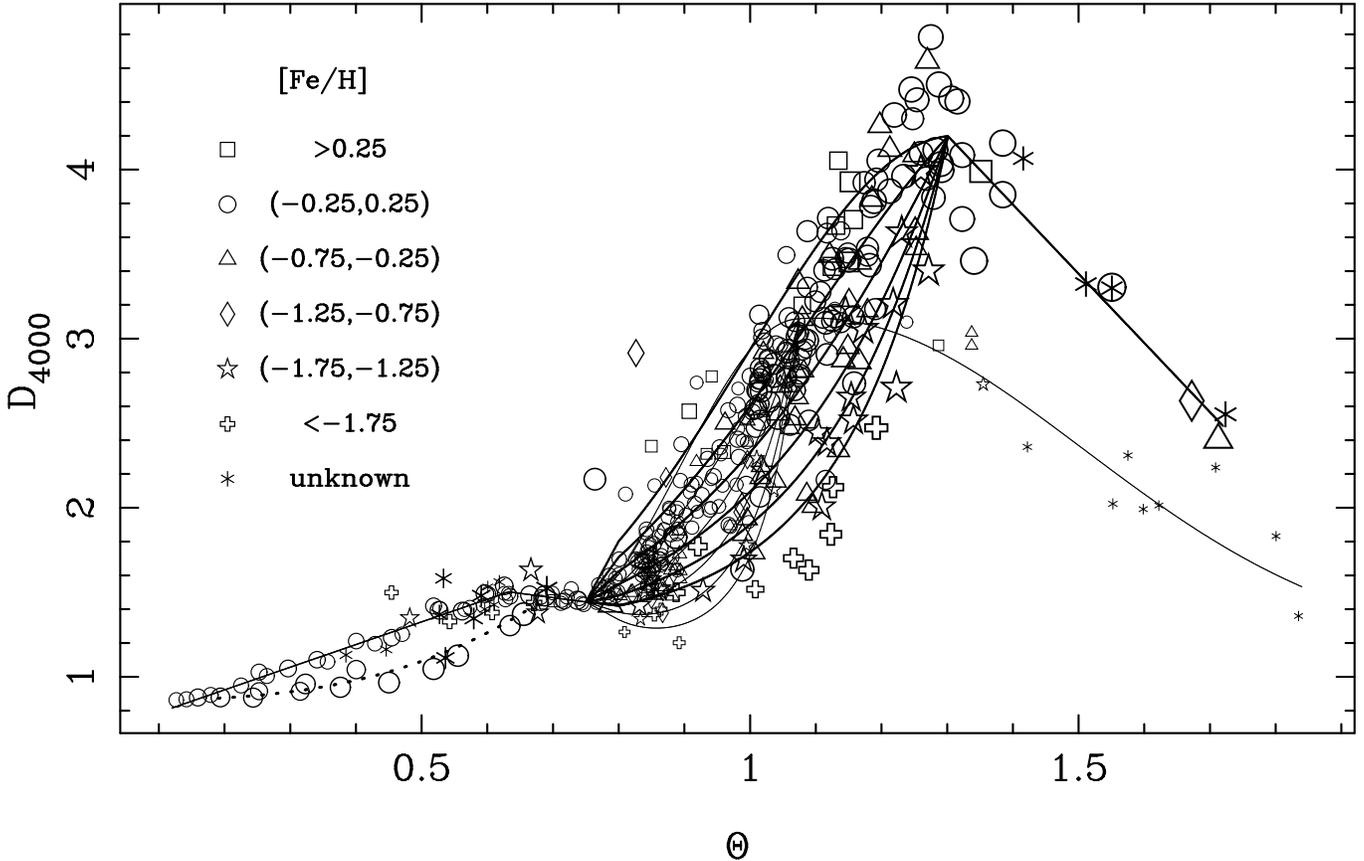}}
  \caption{D$_{4000}$ as a function of $\theta\equiv 5040/T_{\rm eff}$ for the 
sample, together with the derived fitting functions. Stars of different
metallicities are shown with different symbol types, 
with sizes giving an indication of the surface
gravity (in the sense that low-gravity stars, i.e. giants, are plotted with 
larger symbols). Concerning the fitting functions, in the low $\theta$ range, 
the solid line corresponds to dwarf and giant stars, 
whereas the dashed line is 
used for supergiants. For lower temperatures, thick and thin lines refer to
giant and dwarf stars respectively. For each of these groups in  
the mid-temperature range,
the different lines represent the metallicities ${\rm [Fe/H]}=+0.5,0, 
-0.5,-1,-1.5,-2$, from top to bottom.
}   
  \label{fitall}
 \end{center}
\end{figure*}

After a set of trial fits, it was clear that temperature is the main parameter
governing the break. Unfortunately, the behaviour of the D$_{4000}$ could not
be reproduced by a unique polynomial function in the whole temperature range
spanned by the library, forcing us to divide the temperature
interval into several regimes. The derived composite fitting function is shown
in Fig.~\ref{fitall}. In Table~\ref{tablefit} we list the corresponding
coefficients and errors, together with the typical error of the $N$ stars used
in each interval ($\sigma_{\rm typ}^2=N/\sum_{i=1}^n \sigma_i^{-2}$), the
unbiased residual variance around the fit ($\sigma_{\rm std}^2$) and the
determination coefficient ($r^2$).


\begin{table}
\caption{Parameters of the empirical fitting functions in each temperature and 
gravity range}
\label{tablefit}
\begin{center}
\begin{tabular}{@{}l@{}c@{}rcl|l@{}} \hline\hline
\multicolumn{2}{@{}l}{\raisebox{0.ex}[2.5ex][0.ex]{\bf Hot stars}} &
\multicolumn{3}{c}{$0.12<\theta<0.63$} & 
$3<\log g<4.5$ \\ \hline
\multicolumn{4}{@{}l}{polynomial fit} & & $N=39$\\
$c_0$ & : & 0.6548 & $\pm$ & 0.0222 &  $\sigma_{\rm typ}=0.041$ \\
$\theta$ & : & 1.340 & $\pm$ & 0.049 & $\sigma_{\rm std}=0.046$ \\
 & & & & & $r^2=0.96$ \\
\hline\hline
\multicolumn{2}{@{}l}{\raisebox{0.ex}[2.5ex][0.ex]{\bf Warm stars}} &
\multicolumn{3}{c}{$0.63<\theta<0.75$} & 
$3<\log g<4.5$ \\ \hline
\multicolumn{4}{@{}l}{polynomial fit} & & $N=23$\\
$c_0$ & : & 1.823 & $\pm$ & 0.140 & $\sigma_{\rm typ}=0.051$ \\
$\theta$ & : & $-$0.5068 & $\pm$ & 0.2030 & $\sigma_{\rm std}=0.034$ \\
 & & & & & $r^2=0.35$ \\ 
\hline\hline
\multicolumn{2}{@{}l}{\raisebox{0.ex}[2.5ex][0.ex]{\bf Hot supergiants}} &
\multicolumn{3}{l}{$0.19<\theta<0.69$} & 
$0<\log g<3$ \\ \hline
\multicolumn{4}{@{}l}{polynomial fit} & & $N=18$\\
$c_0$ & : & 0.8613 & $\pm$ & 0.0328 &  $\sigma_{\rm typ}=0.041$ \\
$\theta^3$ & : & 1.849 & $\pm$ & 0.211 & $\sigma_{\rm std}=0.079$ \\
 & & & & & $r^2=0.87$ \\
\hline\hline
\multicolumn{2}{@{}l}{\raisebox{0.ex}[2.5ex][0.ex]{\bf Cool dwarfs}} &
\multicolumn{3}{c}{$0.75<\theta<1.08$} &
$3<\log g<5.1$ \\ \hline
\multicolumn{4}{@{}l}{exponential fit (${\rm const.}=0.9$)} & & $N=161$ \\
$c_0$ & : & $-$8.154 & $\pm$ & 2.007 & $\sigma_{\rm typ}=0.068$ \\
$\theta$ & : & 13.45 & $\pm$ & 4.28 & $\sigma_{\rm std}=0.178$ \\
\z & : & $-$17.06 & $\pm$ & 7.28 & $r^2=0.88$ \\
$\theta\ $\z & : & 38.12 & $\pm$ & 15.80 & \\
$\theta^2$ & : & $-$4.769 & $\pm$ & 2.269 & \\
$\theta^2\ $\z & : & $-$20.69 & $\pm$ & 8.52 & \\
\hline\hline
\multicolumn{2}{@{}l}{\raisebox{0.ex}[2.5ex][0.ex]{\bf Cool giants}} &
\multicolumn{3}{c}{$0.75<\theta<1.30$} &
$0<\log g<3.5$ \\ \hline
\multicolumn{4}{@{}l}{exponential fit (${\rm const.}=0.9$)} & & $N=176$ \\
$c_0$ & : & $-$5.665 & $\pm$ & 1.563 & $\sigma_{\rm typ}=0.081$ \\
$\theta$ & : & 9.279 & $\pm$ & 2.766 & $\sigma_{\rm std}=0.261$ \\
\z & : & $-$3.273 & $\pm$ & 2.274 & $r^2=0.85$ \\
$\theta\ $\z & : & 7.322 & $\pm$ & 4.266 & \\
$\theta^2$ & : & $-$3.080 & $\pm$ & 1.218 & \\
$\theta^2\ $\z & : & $-$3.694 & $\pm$ & 1.986 & \\
\hline\hline
\multicolumn{2}{@{}l}{\raisebox{0.ex}[2.5ex][0.ex]{\bf Cold dwarfs}} &
\multicolumn{3}{c}{$1.08<\theta<1.83$} &
$4.5<\log g<5.2$ \\ \hline
\multicolumn{4}{@{}l}{exponential fit (${\rm const.}=1.15$)} & & $N=15$ \\
$c_0$ & : & $-$2.908 & $\pm$ & 2.949 & $\sigma_{\rm typ}=0.078$ \\
$\theta$ & : & 6.535 & $\pm$ & 4.243 & $\sigma_{\rm std}=0.194$ \\
$\theta^2$ & : & $-$2.976 & $\pm$ & 1.506 & $r^2=0.83$ \\
\hline\hline
\multicolumn{2}{@{}l}{\raisebox{0.ex}[2.5ex][0.ex]{\bf Cold giants}} &
\multicolumn{3}{c}{$1.30<\theta<1.72$} &
$0<\log g<1.2$ \\ \hline
\multicolumn{4}{@{}l}{polynomial fit} & & $N=15$ \\
$c_0$ & : & 9.525 & $\pm$ & 0.742 &  $\sigma_{\rm typ}=0.083$ \\
$\theta$ & : & $-$4.094 & $\pm$ & 0.486 & $\sigma_{\rm std}=0.211$ \\
 & & & & & $r^2=0.92$ \\
\hline
\end{tabular}
\end{center}
\end{table}

In the high temperature regime ($\theta\leq0.75$, $T_{\rm eff}\geq6700$ K) a
dichotomic behaviour for dwarfs and giants on one side, and supergiants on the
other, is clearly apparent. Therefore we derived different fitting functions
for each gravity range. For the first group the amplitude of the break is
quite constant and only the linear term in $\theta$ is statistically
significant (note that we subdivide this range in two intervals to achieve a
better fit). The independence on metallicity is naturally expected (see
section~2) but note that an important fraction of the stars in this range
either lack of a [Fe/H] estimation or are restricted to the solar value.

\begin{figure}
 \begin{center}
  \resizebox{\hsize}{!}{\includegraphics[bb= 111 200 500 656,angle=0]{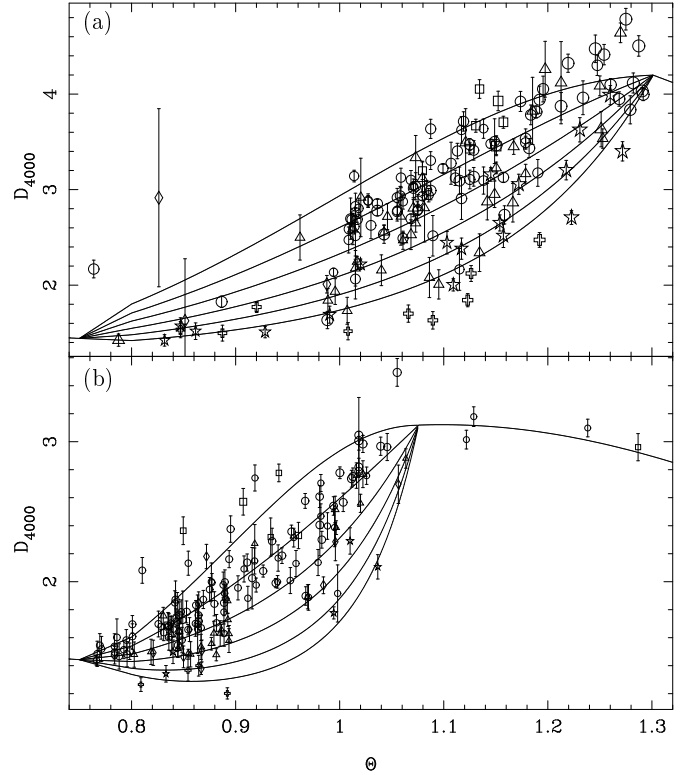}}
  \caption{Details of the fitting functions in the mid-temperature range for
a) giant ($\log g<3.5$) and b) dwarf ($\log g>3$) stars. See
caption to Fig~\ref{fitall}. Error bars for the D$_{4000}$ measurements are 
also shown}
  \label{gigyena}
 \end{center}
\end{figure}

The behaviour of the cool stars
($0.75\leq\theta\leq1.3$, $3900 {\rm K}\leq T_{\rm eff}\leq 6700$ K) is more
complex and [Fe/H] terms are clearly needed. On the other hand, no gravity
term is significant. However, whilst
for the giant stars D$_{4000}$ increases with $\theta$ all the way up to
$\theta\approx1.3$, for higher gravities it reaches a maximum at
$\theta\approx1.1$ and then levels off. Furthermore, separate fits for dwarfs
and giants in this $T_{\rm eff}$ range (with a gravity cutoff around $3-3.5$)
yield residual variances that are significantly smaller than the variance
from a single fit. Hence, we have derived different fitting functions
for dwarfs and giants. This dichotomic behaviour of the break is not
surprising since its strength is quite dependent on the depth of the CN bands
(Fig.~1) which also shows a similar behaviour (G93) due to the onset of the
dredge-up processes at the bottom of the giant branch. In Fig.~\ref{gigyena} 
we show in detail the fitting functions derived for each gravity group in this
temperature range. 

Concerning the cold stars ($\theta\geq1.3$, $T_{\rm eff}\leq3900$ K), the
difference between giants and dwarfs is quite evident and two fitting
functions have been derived (see also section~2). Again, the metallicity terms
are not significant, although this may be, at least in part, due to the
paucity of input metallicities in this range.  It must be noted that the
different fitting functions have been constructed with the constrain of
allowing for a smooth transition in the predicted D$_{4000}$ indices among the
different $T_{\rm eff}$ and gravity ranges.

\begin{figure}
 \begin{center}
  \resizebox{\hsize}{!}{\includegraphics[bb= 111 5 500 781,angle=0]%
{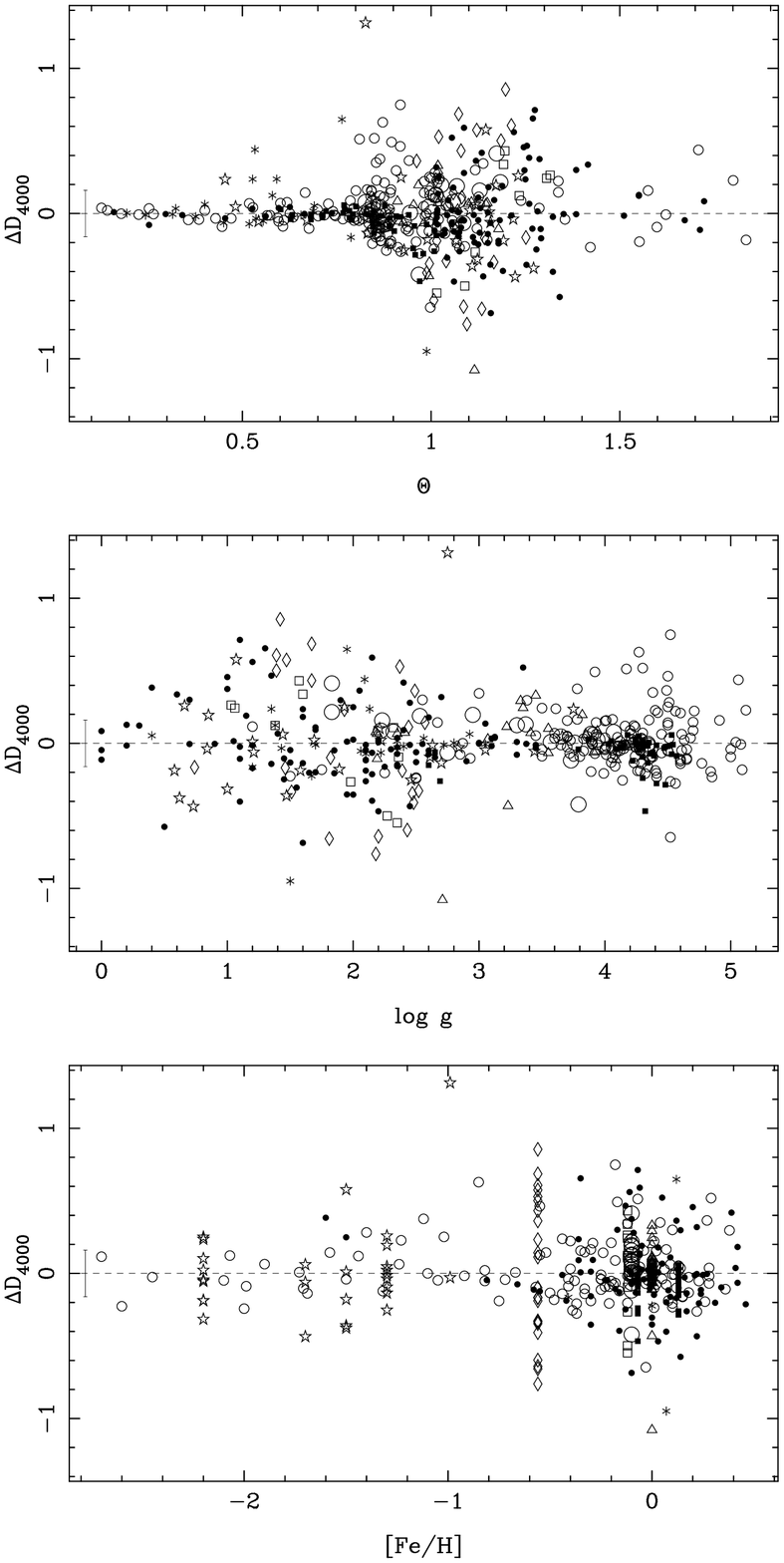}}
  \caption{Residuals of the derived fitting functions (observed minus 
predicted) 
against the three input stellar parameters. Symbol types are the same as in
Fig~\ref{plotspt}. The length of the error bar is twice the unbiased residual 
standard deviation.}
  \label{figresiduals}
 \end{center}
\end{figure}

In Fig.~\ref{figresiduals} we plot the residuals from the fits as a function
of effective temperature, metallicity and gravity. Note that no trends are
apparent with any of these parameters. We have also checked for systematic
residuals within any of the star clusters. Except for an unexplained negative
offset for the Coma stars ($\Delta{\rm D}_{4000}=0.09$, not due to an error in
the adopted metallicity), no systematic offsets have been found. 
For the 420 stars used in the fit, we derive an unbiased residual standard
deviation $\sigma_{\rm std}=0.160$. This must be compared with the typical
error in the D$_{4000}$, $\sigma_{\rm typ}=0.064$. Therefore, the residuals
are, in the mean, a 2.5 factor larger than what should be expected solely from
measurement errors. Since we are quite confident that this latter errors are
realistic (see Section~\ref{secerr}), and although some scatter may arise from
the fact than the fitting functions are not able to reproduce completely the
complex behaviour of the D$_{4000}$, most of the extra scatter must arise form
uncertainties in the input atmospheric parameters. For example, the residual
D$_{4000}$ scatter of 0.248 for the cool giants (at $\theta = 1.0$ and 
${\rm [Fe/H]}
= 0.0$) can be fully explained by the combined effect of a 166 K uncertainty
in $T_{\rm eff}$ and a 0.29 dex error in [Fe/H], both consistent
with the typical errors found by Soubiran,
Katz \& Cayrel (1998) when comparing atmospheric parameters from 
the literature.  
Another quantitative measurement of the quality of the present fitting
functions is the determination coefficient for the whole sample $r^2=0.96$.
This indicates that a 96\% of the original variation of the break in the sample
is explained by the derived fitting functions.

Since the goal of this work is to predict reliable D$_{4000}$ indices for any
given combination of input atmospheric parameters, 
we have investigated, using the
covariance matrices of the fits, the random errors in such predictions. These
errors are given in
Table~\ref{finerr} for some representative sets of input parameters. Note
that, as it should be expected, the uncertainties are smaller
for near-solar metallicities. Interestingly, although the library does not 
include a high number of 0--B stars, 
the predicted indices at the 
hot end of the star sample are rather reliable.

\begin{table}
\caption{Uncertainties in the predicted D$_{4000}$ values for some indicative
sets of stellar parameters. For the cool stars ($T_{\rm eff}\leq6000 {\rm K}$) 
the term {\it giants\/} refers also to supergiant stars.}
\label{finerr}
\begin{center}
\begin{tabular}{rrcc} \hline\hline
\multicolumn{1}{c}{\raisebox{0.ex}[2.5ex][1.0ex]{$T_{\rm eff}$}} & 
\multicolumn{1}{r}{[Fe/H]} &
\multicolumn{1}{c}{dwarfs--giants} & \multicolumn{1}{c}{supergiants} \\
\hline
\raisebox{0.ex}[3.0ex][0.ex]{30000} & & 0.015 & 0.032 \\
\raisebox{0.ex}[3.0ex][0.ex]{12000} & & 0.008 & 0.023 \\
\raisebox{0.ex}[3.0ex][0.ex]{7500} & & 0.010 & 0.044 \\
\hline
\multicolumn{1}{c}{\raisebox{0.ex}[4.ex][1.0ex]{\mbox{ }}} & &
\multicolumn{1}{c}{dwarfs} & \multicolumn{1}{c}{giants} \\
\hline
\raisebox{0.ex}[3.0ex][0.ex]{6000} & 0.5 & 0.045 & 0.156 \\
6000 & 0.0 & 0.024 & 0.099 \\
6000 & $-$1.0 & 0.031 & 0.075 \\
6000 & $-$2.0 & 0.043 & 0.105 \\
\raisebox{0.ex}[3.0ex][0.ex]{5000} & 0.5 & 0.103 & 0.079 \\
5000 & 0.0 & 0.044 & 0.043 \\
5000 & $-$1.0 & 0.082 & 0.053 \\
5000 & $-$2.0 & 0.119 & 0.073 \\
\raisebox{0.ex}[3.0ex][0.ex]{4000} & 0.5 & & 0.167 \\ 
4000 & 0.0 & 0.125 & 0.088 \\
4000 & $-$1.0 & & 0.173 \\
4000 & $-$2.0 & & 0.322 \\
\raisebox{0.ex}[3.0ex][0.ex]{3200} & & 0.094 & 0.082 \\
\hline
\end{tabular}
\end{center}
\end{table}


\section{Summary}

We have derived a set of empirical fitting functions describing the
behaviour of the break at $\lambda 4000$ \AA\ in terms of the
atmospheric stellar parameters: effective temperature, metallicity and
surface gravity. This calibration can be easily incorporated into
stellar population models to provide accurate predictions of the
D$_{4000}$ for composite systems. In a forthcoming paper we will
analyze the measurements of the break in old stellar populations at
the light of the predictions of such models.  Considering the volume
covered by the employed stellar library (Lick/IDS $+$ Pickles' hot
subsample) in the stellar parameter space, the derived fitting
functions suit the requirements to provide accurate D$_{4000}$
predictions for populations with ages larger than about 0.1~Gyr, and
$-1\le {\rm [Fe/H]} \le +0.5$~dex (see also W94). Note however that,
since the break could be contaminated by nebular emission in galaxies
with ongoing star formation, it is necessary to include this effect in
the population modelling in order to predict reliable D$_{4000}$
indices for such stellar populations.
It should also be noted that the 
applicability
of the derived fitting functions is safe as far as the abundance ratios of the
library stars reflect those in the modeled stellar populations.

In order to facilitate the usage of the present D$_{4000}$ fitting functions 
we have written a FORTRAN subroutine, available from:\\
{\small\tt http://www.ucm.es/info/Astrof/D4000/D4000.html}.\\ 
This routine computes the value of the D$_{4000}$ as a function of the input
stellar parameters $T_{\rm eff}$, [Fe/H] and $\log g$. The code performs
smooth interpolations among temperature and gravity ranges, providing also an
estimate of the error in the predicted index,

\begin{acknowledgements}
We thank Jes\'{u}s Gallego (runs 3 and~5) and Jaime Zamorano (run~3) for
carrying out the observations of some stars of the sample. We also thank
Emilios Harlaftis, Mart\'{\i}n Guerrero and Reynier Peletier, the support
astronomers who performed the service time observations of runs 7, 10 and~12,
respectively. We are grateful to the anonymous referee for useful suggestions.
 The JKT, INT and WHT are operated on the island of La Palma by
the Royal Greenwich Observatory at the Observatorio del Roque de los Muchachos
of the Instituto de Astrof\'{\i}sica de Canarias. The Calar Alto Observatory is
operated jointly by the Max-Planck-Institute f\"{u}r Astronomie, Heidelberg,
and the Spanish Comisi\'{o}n Nacional de Astronom\'{\i}a. This research has
made use of the Simbad database (operated at CDS, Strasbourg, France), the
NASA's Astrophysics Data System Article Service, and the Hipparcos
Catalogue. This work was supported by the Spanish 'Programa Sectorial de
Promoci\'{o}n del Conocimiento' under grant \mbox{No.~PB96-610}.
\end{acknowledgements}

{}

\end{document}